\documentclass[10pt,twocolumn,letterpaper]{article}

\usepackage[pagenumbers]{cvpr} 
\definecolor{cvprblue}{rgb}{0.21,0.49,0.74}
\usepackage[pagebackref,breaklinks,colorlinks,allcolors=cvprblue]{hyperref}

\usepackage{amsmath}
\usepackage{multirow}
\usepackage{multicol}
\usepackage{tikz}
\usepackage[ruled,lined,linesnumbered]{algorithm2e}
\usepackage{enumitem}
\usepackage[dvipsnames, table]{xcolor}

\newcommand{\goldtext}[1]{{\textcolor{YellowOrange}{\textbf{#1}}}\xspace}
\newcommand{\silvertext}[1]{{\textcolor{Gray}{\textbf{#1}}}\xspace}

\newcommand{\gold}{}
\newcommand{\silver}{}

\newcommand*\circled[2]{\tikz[baseline=(char.base)]{
            \node[shape=circle,fill=black,inner sep=1pt] (char) {\textcolor{#1}{{\footnotesize #2}}};}}

\newcommand{\redright}{\textcolor{red}{$\rightarrow$}\xspace}

\ifx\figurename\undefined \def\figurename{Figure}\fi
\renewcommand{\figurename}{Fig.}
\renewcommand{\paragraph}[1]{\textbf{#1} }

\newcommand{\Sect}[1]{Sec.~\ref{#1}}
\newcommand{\Fig}[1]{Fig.~\ref{#1}}
\newcommand{\Tbl}[1]{Tbl.~\ref{#1}}

\newcommand{\specialcell}[2][c]{\begin{tabular}[#1]{@{}c@{}}#2\end{tabular}}

\def\Item{{\noindent\textbullet\hspace{0.5em}}}

\newcommand{\proj}{\textsc{TimeRipple}\xspace}
\newcommand{\mode}[1]{\underline{\textsc{#1}}\xspace}

\graphicspath{{figs/}}

\def\paperID{12440} 
\def\confName{CVPR}
\def\confYear{2026}

\title{\proj: Accelerating vDiTs by Understanding the Spatio-Temporal Correlations in Latent Space}

\author{
  Wenxuan Miao$^{1}$,
  Yulin Sun$^{1}$,
  Aiyue Chen$^{3}$,
  Jing Lin$^{3}$,
  Yiwu Yao$^{3}$,\\
  Yiming Gan$^{4}$,
  Jieru Zhao$^{1}$,
  Jingwen Leng$^{1, 2}$,
  Mingyi Guo$^{1, 2}$,
  Yu Feng$^{1,2,\dagger}$\thanks{$\dagger$ Corresponding author.}\\
  $^{1}$Shanghai Jiao Tong University, \quad
  $^{2}$Shanghai Qizhi Institute,\quad \\
  $^{3}$Huawei Technologies Co.,Ltd,\quad
  $^{4}$Institute of Computing Technology, Chinese Academy of Science
}

\begin{document}
\maketitle
\begin{abstract}

The recent surge in video generation has shown the growing demand for high-quality video synthesis using large vision models. 
Existing video generation models are predominantly based on the video diffusion transformer (vDiT), however, they suffer from substantial inference delay due to self-attention. 
While prior studies have focused on reducing redundant computations in self-attention, they often overlook the inherent spatio-temporal correlations in video streams and directly leverage sparsity patterns from large language models to reduce attention computations.

In this work, we take a principled approach to accelerate self-attention in vDiTs by leveraging the spatio-temporal correlations in the latent space. 
We show that the attention patterns within vDiT are primarily due to the dominant spatial and temporal correlations at the token channel level.
Based on this insight, we propose a lightweight and adaptive reuse strategy that approximates attention computations by reusing partial attention scores of spatially or temporally correlated tokens along individual channels.
We demonstrate that our method achieves significantly higher computational savings (85\%) compared to state-of-the-art techniques over 4 vDiTs, while preserving almost identical video quality ($<$0.06\% loss on VBench).

\end{abstract}

\section{Introduction}
\label{sec:intro}

Virtually, no word has drawn more attention than \textit{Sora} in 2024. 
In this year alone, more than 50 research institutions worldwide have launched their own generative frameworks for video creation~\cite{kling, sora, hunyuan, veo2, opensora, lin2024open, hong2022cogvideo, singer2022make, mlgen2, xu2024easyanimate, seawead2025seaweed, zeng2024make, genmo2024mochi, wan2025}. 
People have witnessed the potential of video generation using large vision models across many facets of daily life, including video editing~\cite{team2025vidi}, content recreation~\cite{liu2024rec}, and more.
However, as the demand for high-quality, high-resolution video continues to grow, these models need to improve their computational efficiency to enable commercial deployment.
Naturally, addressing their efficiency bottleneck is an important issue.

Despite numerous frameworks, existing large vision models generally adopt the same inference paradigm, video diffusion transformer (vDiT).  
Currently, there are two main factors that affect the inference latency of vDiT models: lengthy \textit{denoising steps} and the compute-intensive \textit{self-attention} modules. 
To address these limitations, existing efforts to accelerate vDiT fall into two categories: reducing the number of denoising steps~\cite{li2023autodiffusion, salimans2022progressive, yin2024one, gu2023boot}, or eliminating redundant computations in self-attention~\cite{xia2025training, yuan2024ditfastattn, ding2025efficient, xi2025sparse}. 
This paper focuses on reducing the self-attention compute.

\begin{figure*}
    \centering
    \includegraphics[width=\textwidth]{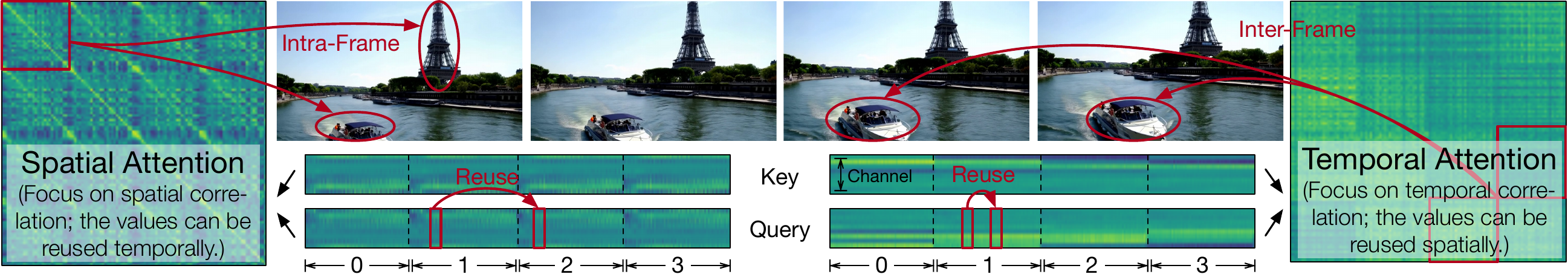}
    \caption{An illustration of spatial and temporal patterns in one head of multi-head attention maps. Due to the space limit, only 4 frames are shown here. The attention patterns are determined by the spatial and temporal correlations of the key and query.
    Spatial-dominated attention (on the left) primarily focuses on the spatial correlations within a frame; thus, the values between two frames are similar and can be reused.
    Temporal-dominated attention (on the right) primarily focused on the temporal correlations across frames; thus, the values within a frame are similar and can be reused.
    }
    \label{fig:teaser}
\end{figure*}

Existing approaches~\cite{xia2025training, yuan2024ditfastattn, ding2025efficient, xi2025sparse} that accelerate the self-attention in vDiTs are mainly inspired by techniques from accelerating large language models (LLMs).
These techniques exploit the sparsity patterns of the attention maps in self-attention and skip insignificant computations.
However, we argue that directly transplanting techniques from LLMs often overlooks the unique properties of video data, the inherent spatio-temporal correlation. 
Long before the era of LLMs, studies had recognized the spatial and temporal correlations within video sequences and leveraged these properties to accelerate various tasks, such as video compression~\cite{bhaskaran1997image, ma2019image}, real-time object tracking~\cite{wang2020towards, feng2022real}, etc.
For instance, compression standards like MPEG-H.265 can achieve over 10$\times$ data reduction without quality loss~\cite{gupta2017modern}.
This shows that significant redundancy exists in video data.

\paragraph{Insight.} 
Motivated by this, we begin with an in-depth analysis of the self-attention patterns within vDiTs.
By dissecting the self-attention computation, we find that tokens along individual channels also exhibit spatial and temporal correlations, i.e., spatially and temporally correlated channels have similar values, similar to those observed in video data.
These correlations at the channel level accumulate together during the attention computation and lead to various patterns in attention maps (\Fig{fig:teaser}).
In other words, \textit{patterns in attention maps are rather the artifacts of the spatial and temporal correlations among tokens in the latent space.}

\paragraph{Idea.}
Building on this insight, we propose a simple yet principled technique to reduce the computation of self-attention, the most compute-intensive operation in vDiTs (\Sect{sec:method:framework}).
Knowing that spatially or temporally correlated tokens within a given channel have a similar value, the partial attention score of a given channel can simply approximate that of the other.
Our idea is to reuse partial attention scores of the spatially or temporally similar tokens to approximate attention computations along individual channels (\Fig{fig:idea}).
\Tbl{tab:overall} shows that leveraging the channel-level correlation leads to higher compute savings than prior studies.

With our reuse-based idea, the next question is how to determine which channel of a given token should be reused.
Our observation is that, under a fixed reuse ratio, the quality of generated videos exhibits a strong sensitivity to denoising steps, rather than to individual prompts (\Fig{fig:threshold_prompt}).
To identify the optimal reuse configuration, we construct an analytical model that captures the correlation between the reuse ratio and the introduced loss at different denoising steps.
Based on this model, our framework adaptively selects reuse thresholds for each denoising step and ensures that our reuse technique introduces a consistent and controllable amount of errors into the final generated videos.

To summarize, this paper offers a new direction for efficient vDiTs via a deep understanding of spatio-temporal correlations in latent space. Our contributions are:

\Item We provide an in-depth analysis of the causes behind the various patterns in attention maps during the diffusion process. We identify that dominant spatio-temporal correlations along token channels govern the final attention pattern.

\Item We propose an adaptive token-reuse method that exploits spatio-temporal patterns along the channel dimension and significantly reduces redundancies in the self-attention.

\Item We demonstrate that our proposed technique achieves the highest end-to-end speedup (up to 2.7$\times$) while maintaining nearly identical generation quality compared to the original models, with much better consistent results ($>$9~dB in PSNR) compared to the state-of-the-art techniques.

\section{Related Work}
\label{sec:related}

\paragraph{Efficient vDiT.}
Early diffusion models~\cite{peebles2023scalable, ho2020denoising, song2020score} require thousands of denoising steps. DDIM~\cite{song2020denoising}, DPM-Solver~\cite{lu2022dpm1, lu2022dpm2}, and flow matching~\cite{lipman2022flow} reduce generation steps via reformulated ODEs or vector fields.

Other methods exploit similarities across denoising steps to accelerate generation, such as PAB~\cite{zhao2024real} and ToCa~\cite{zou2024accelerating}, which reuse intermediate results. Token merging techniques like ToMe~\cite{bolya2023token} reduce attention cost by shortening token sequences. However, these approaches often require non-trivial memory or compute overhead. Our method differs by introducing negligible overhead while achieving higher computational savings via spatio-temporal correlations.

\paragraph{Sparsity in Attention.}
Attention sparsity has been widely explored in LLMs~\cite{child2019generating, xiao2023efficient, zhang2023h2o, jiang2024minference}, but such patterns often fail to generalize to vDiTs~\cite{kong2024hunyuanvideo, yang2024cogvideox, wan2025, lin2024open}. 

Existing vDiT acceleration methods apply fixed or predefined sparse patterns in attention maps, e.g., sliding windows~\cite{yuan2024ditfastattn}, tiling~\cite{ding2025efficient}, or masking~\cite{xi2025sparse}. AdaSpa~\cite{xia2025training} adopts a dynamic reuse strategy to address diversity, yet still focuses solely on the attention map.

In contrast, we analyze the root cause of pattern diversity, i.e., spatio-temporal correlations in latent space, and propose a reuse-based attention strategy.

\section{Methodology}
\label{sec:method}

In this section, we first provide the necessary background on vDiT (\Sect{sec:method:pre}).
We then present an in-depth analysis of the underlying causes of the various patterns observed in self-attention (\Sect{sec:method:why}).
Finally, we introduce a simple token-reuse technique that accelerates vDiTs (\Sect{sec:method:framework}).

\subsection{Preliminary}
\label{sec:method:pre}

\begin{figure*}[t]
    \centering
    \includegraphics[width=0.95\textwidth]{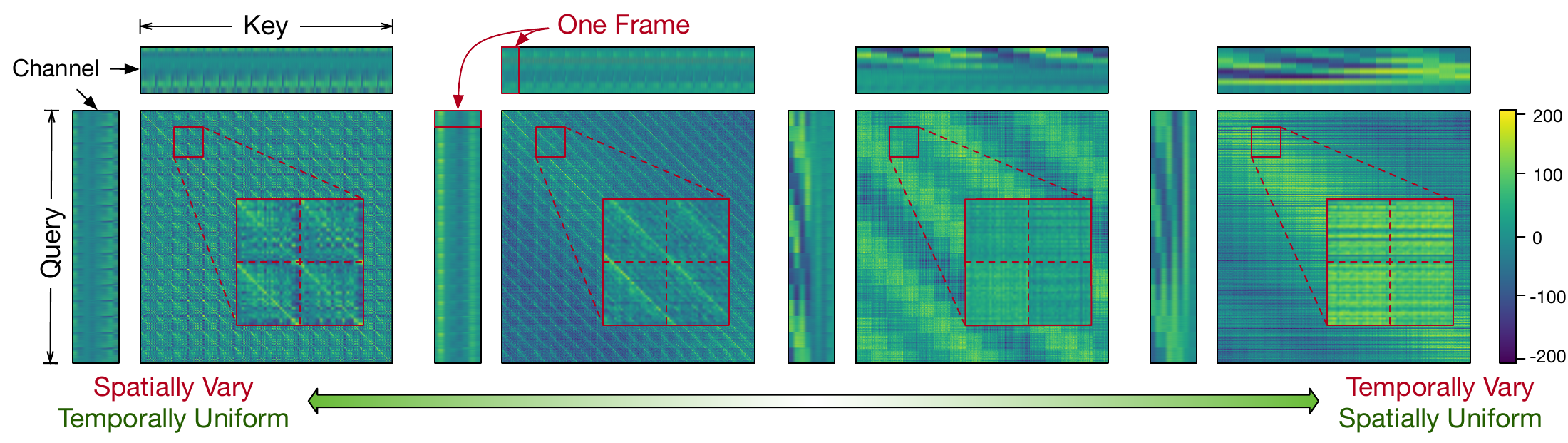}
    \caption{Examples of attention maps with different patterns.
    For visualization, we present only a fraction ($\frac{1}{2}$) of the full attention map along each dimension and zoom in the attention scores of $2\times2$ frames on each attention map.
    On the left, \textit{spatially-varying} attention maps primarily capture spatial information within individual frames.
    Spatially-dominated attention maps often have no significant variations across frames; thus, the values across frames are similar.
    On the right, as temporal-oriented channels dominate, \textit{temporally-varying} attention maps increasingly focus on temporal correlations across frames; thus, the values within a frame are similar.
    The color bar shows the magnitude of attention scores and does not correspond to the values of $Q$ or $K$.
    }
    \label{fig:pattern}
\end{figure*}

\begin{figure}[t]
    \centering
    \includegraphics[width=\columnwidth]{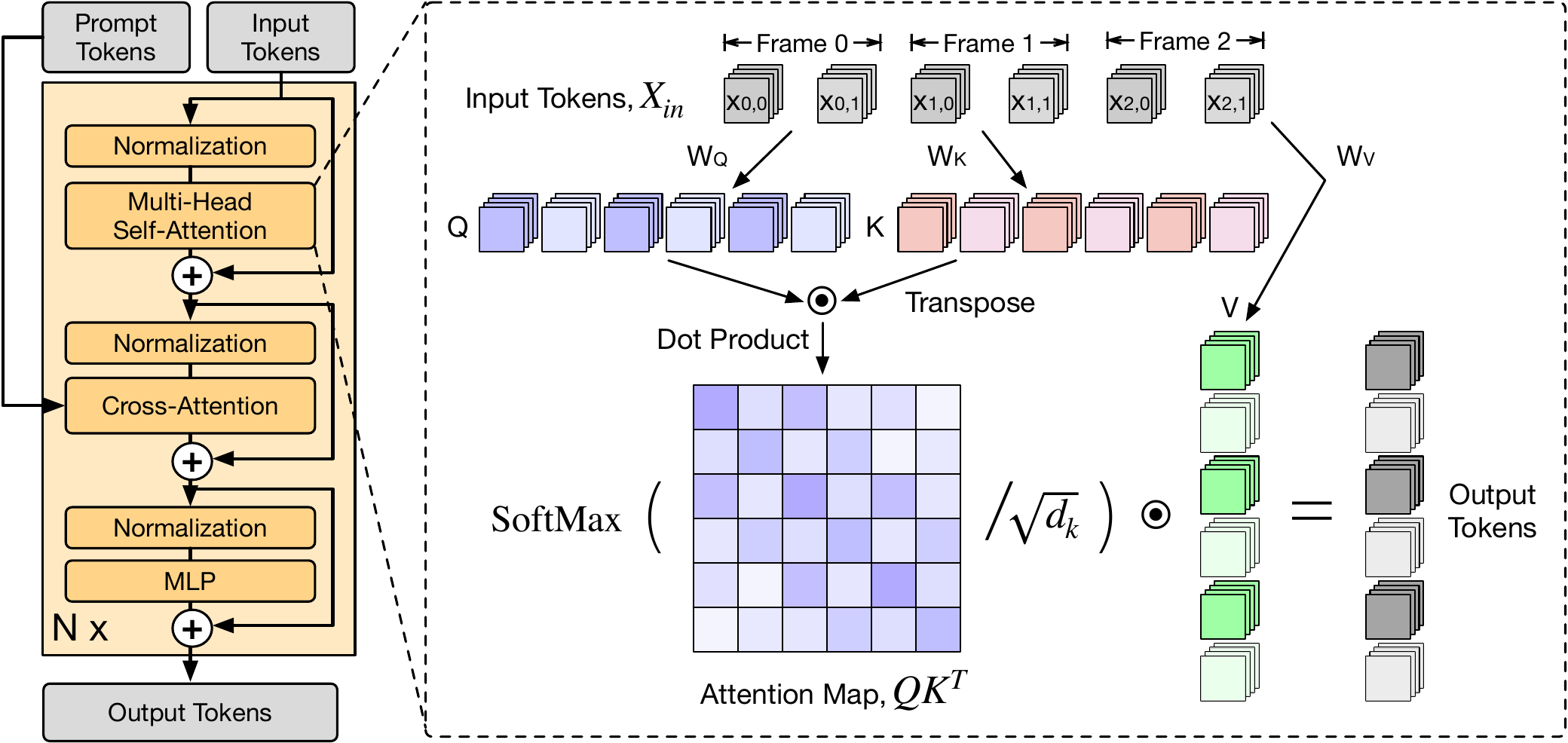}
    \caption{The overview of vDiT architectures. A vDiT consists of multiple blocks. Generally, each block contains a self-attention layer, a cross-attention layer, and a linear layer. }
    \label{fig:vdit}
\end{figure}

\begin{figure}[t]
    \centering
    \includegraphics[width=\columnwidth]{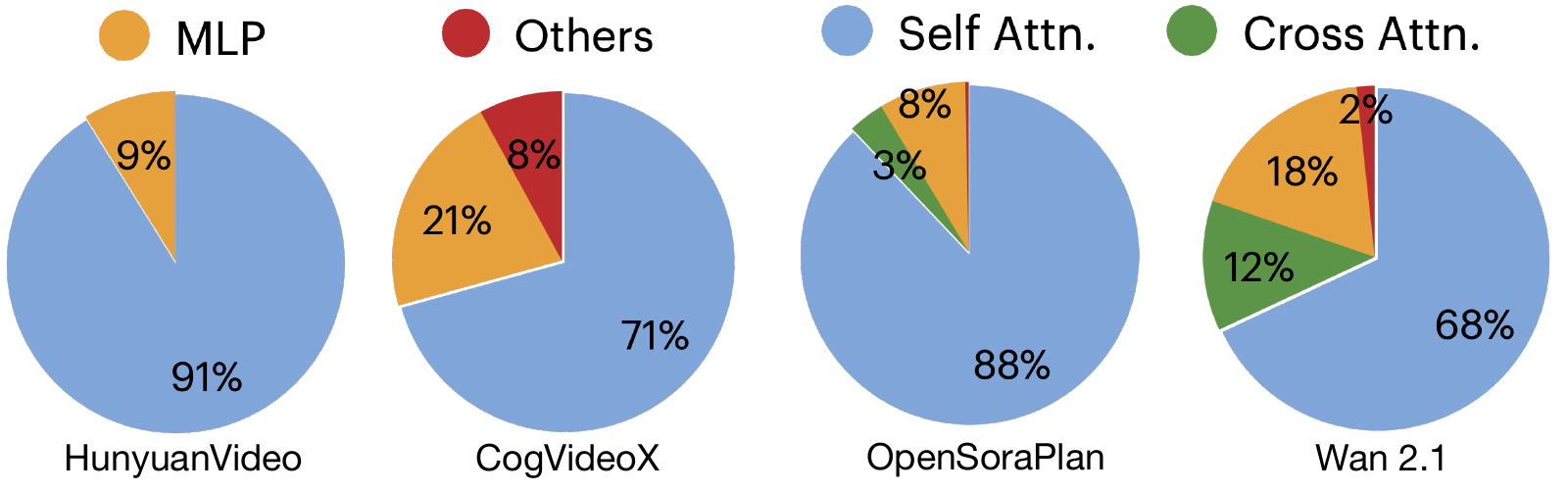}
    \caption{The execution breakdown of four popular vDiT models~\cite{lin2024open, hunyuan, hong2022cogvideo, wan2025} on a single Nvidia H100 (80~GB). 
    The computation of self-attention dominates the execution.
    }
    \label{fig:breakdown}
\end{figure}

\paragraph{Diffusion Transformer.}
Diffusion models generate data from Gaussian noise via a reverse Markov process~\cite{croitoru2023diffusion, yang2023diffusion, ho2020denoising}. 
At each denoising step $t$, the model predicts noise $z'_t$ and removes it from the input $x'_t$, yielding the final output $x'_0 \approx x_0$.
\Fig{fig:vdit} shows a typical vDiT architecture, which encodes input and prompt tokens into $d$-dimensional latent space. 
The model consists of stacked blocks with self-attention, cross-attention, and MLP layers. 
\Fig{fig:breakdown} profiles four popular vDiT models~\cite{lin2024open, hunyuan, hong2022cogvideo, wan2025} reveals that self-attention dominates inference latency, accounting for 78\% of total execution time, on average.

\paragraph{Self-Attention.}
\Fig{fig:vdit} shows that, given token input $X_\text{in} \in \mathbb{R}^{N \times d}$, self-attention computes,
\begin{equation}
    \text{Attention}(Q, K, V) = \text{Softmax}\left(\frac{QK^T}{\sqrt{d_k}}\right)V,
\end{equation}
where $Q, K, V$ are linear projections of $X_\text{in}$ and $A = QK^T \in \mathbb{R}^{N \times N}$ is the attention map.

\paragraph{Position Encoding.}
To preserve spatial-temporal information, all vDiTs apply rotary positional embedding (RoPE)~\cite{lin2024open, hunyuan, hong2022cogvideo, wan2025}, which encodes relative positions,
\begin{equation}
    \text{RoPE}([x, y]) = (\cos\theta \cdot x - \sin\theta \cdot y,\ \sin\theta \cdot x + \cos\theta \cdot y),
\end{equation}
where $\theta_{p,i} = \varphi / 10000^{2i/d}$. Importantly, the $d$ channels are often partitioned by semantic roles: initial channels encode temporal information, while others capture spatial positions along $x$ and $y$ axes. 
This structured encoding causes different channel groups to exhibit distinct spatio-temporal behaviors, which we exploit in our reuse strategy.

\subsection{Spatio-Temporal Correlations}
\label{sec:method:why}

\begin{figure*}[t]
    \centering
    \includegraphics[width=\textwidth]{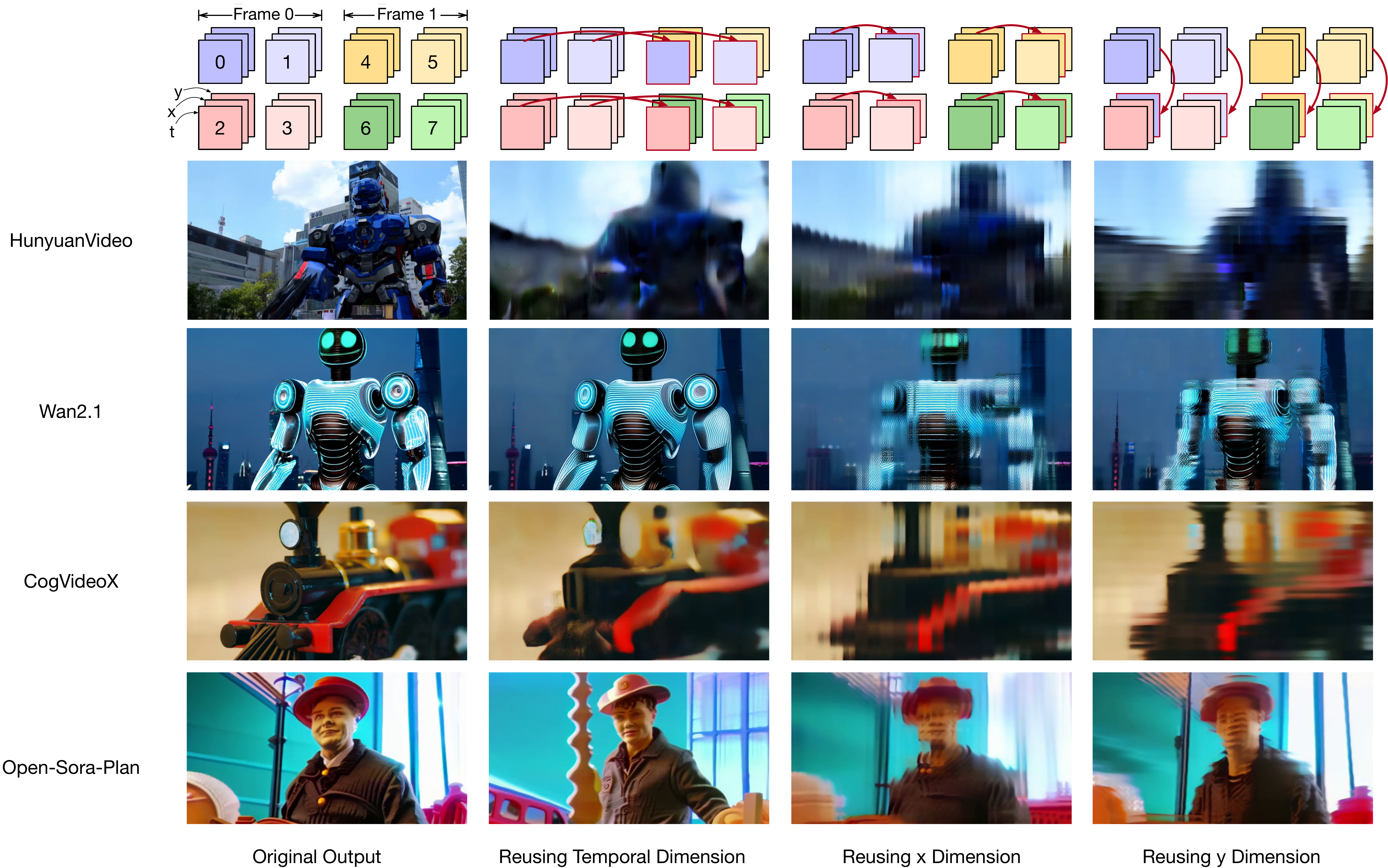}
    \caption{The effects of how different channel groups govern the final generation quality. Here, we maliciously reuse \textit{all tokens} at a particular channel group. The upper part shows how we reuse different channel groups in red arrows. We only show three channels to represent time, the x-axis, and the y-axis channel groups. The lower part shows result figures; the overall videos have similar effects (see supplementary). Note that, our reuse strategy will \textit{not} introduce artifacts as we show here.}
    \label{fig:reuse_effect}
\end{figure*}

\paragraph{Patterns in Attention Maps.}
Prior works~\cite{xia2025training, yuan2024ditfastattn, ding2025efficient, xi2025sparse} have pointed out that vDiTs often exhibit distinctive patterns in their attention maps, as shown in \Fig{fig:pattern}.
Some~\cite{xi2025sparse} have hypothesized that different attention maps are responsible for capturing different correlations.
For example, attention maps with spatially-varying patterns (on the left of \Fig{fig:pattern}) often capture spatial significances within individual frames.
In contrast, attention maps showing diagonal structures with less intra-frame variation (on the right of \Fig{fig:pattern}) correspond to temporal correlations across frames.

We further verify this hypothesis by dissecting the attention computation on HunyuanVideo~\cite{hunyuan}.
Our key conclusion is that \textit{different attention patterns are formed by the synergy of the important channels in both the query ($Q$) and key ($K$)}.
In \Fig{fig:pattern}, we present four attention maps, transitioning from spatially-varying to temporally-varying patterns.
For each pattern, we also show 8 channels that contribute most to the final attention scores.
The red bounding boxes (bboxes) on $Q$ and $K$ highlight individual frames, with colors representing different token values.

Our results show that the overall attention map is considered as ``spatially dominated'' when both $Q$ and $K$ exhibit spatial correlations, i.e., the same pattern is repeated across frames.
On the other hand, both $Q$ and $K$ on the right of \Fig{fig:pattern} show the temporal correlations, i.e., the average value of different frames varies while the value within a frame stays relatively the same, thus, this attention map shows temporal patterns.

\paragraph{Token Channels.}
We also show that different channels often govern different types of information.
As introduced in \Sect{sec:method:pre}, all vDiTs use RoPE to encode positional information into different channels.
In particular, vDiT models split the entire channel dimension into three separate groups: time, x-direction, and y-direction.
For example, in HunyuanVideo, the first 16 channels encode temporal information, the next 56 channels correspond to spatial information along the x-direction, and the final 56 channels capture spatial information along the y-direction.

\begin{figure*}[t]
    \centering
    \includegraphics[width=0.95\textwidth]{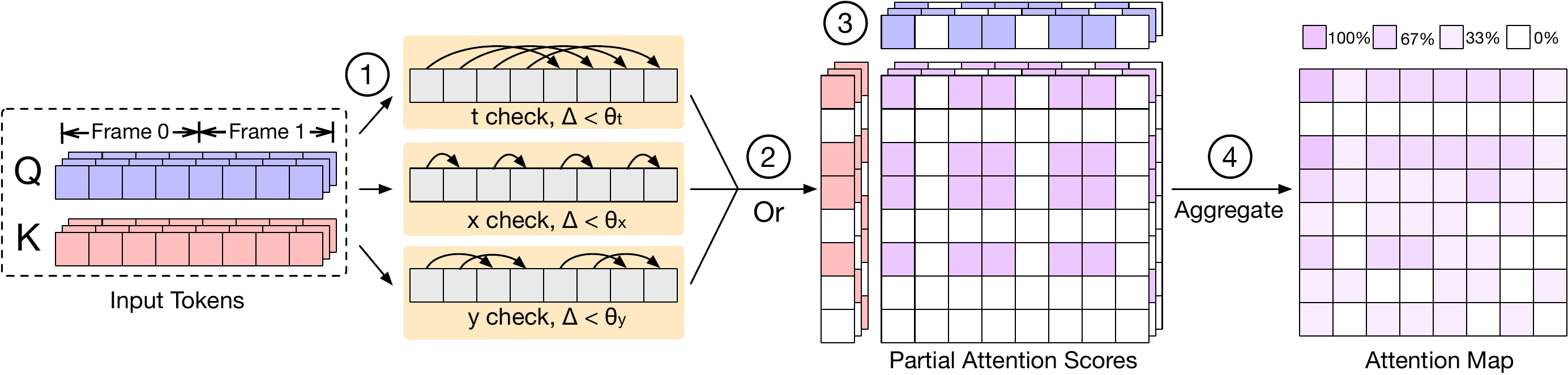}
    \caption{The overview of our reuse method. 
    1) Both $Q$ and $K$ undergo three checks to identify tokens that can reuse the partial attention scores of previous tokens. 
    2) We next collect the reuse patterns.
    3) Then, we sparsely compute the partial attention scores along each channel for non-reuse tokens. 
    4) Lastly, the attention map is obtained by aggregating all partial scores.
    Different colors in the attention map show the percentages of partial attention scores obtained by computation.
    }
    \label{fig:idea}
\end{figure*}

\Fig{fig:reuse_effect} shows how \textit{maliciously} manipulating different channel groups affects the final video quality.
Specifically, we manipulate one particular channel group by forcing the second frame/row/token in every consecutive pair to reuse a specific channel group from the first one across all steps.

The upper part of \Fig{fig:reuse_effect} shows a toy example of the effects after reusing different channel groups.
For illustration purposes, we only use eight input tokens (two frames with $2\times2$ tokens) here.
In \Fig{fig:reuse_effect}, we reuse specific channel groups for every two adjacent frames across all denoising steps.
For instance, to reuse the temporal channels, we only reuse the channels that capture the temporal information while keeping the other two channel groups untouched.

Due to space limitations, we only pick a single image from all four vDiT models~\cite{lin2024open, hunyuan, hong2022cogvideo, wan2025} in \Fig{fig:reuse_effect}.
However, the overall trend holds across the entire video.
Reusing the time-related channels introduces temporal distortions.
In contrast, reusing channels corresponding to the x- or y-direction produces stripe-like artifacts aligned with the corresponding axis.
Overall, all used frames are much blurrier than the original ones because the self-attention mechanism integrates global contributions from of all tokens.

\subsection{Our Acceleration Technique}
\label{sec:method:framework}

\paragraph{The Scope of This Work.}
While numerous acceleration techniques exist, such as skipping denoising steps~\cite{li2023autodiffusion}, reusing intermediate attention results~\cite{zhao2024real}, and masking attention computations~\cite{xi2025sparse}, our work primarily focuses on exploring more effective directions to reduce attention computations, rather than aiming for the optimal acceleration of vDiT as a whole.
\Sect{sec:eva:perf} shows that our technique can theoretically achieve the highest savings on attention computation, and it can be integrated with conventional masking techniques~\cite{xi2025sparse}.
That said, how to incorporate our reuse mechanism with the off-the-shelf attention acceleration methods, e.g., flash-attention~\cite{dao2022flashattention}, is left for future work.

\begin{figure}[t]
\centering
\begin{minipage}[t]{0.48\columnwidth}
  \centering
  \includegraphics[width=\columnwidth]{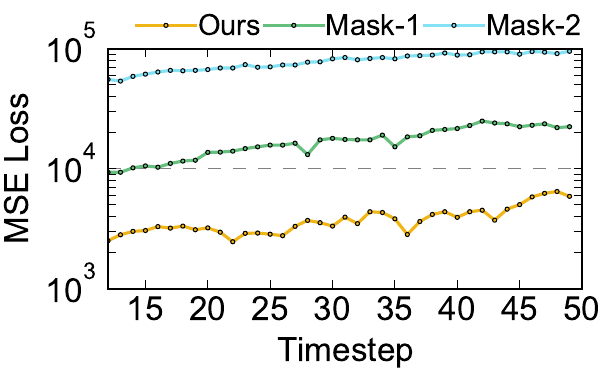}
  \caption{MSE comparison between our reusing method and two masking baselines. Our loss is much lower.}
  \label{fig:reuse_cmp}
\end{minipage}
\hspace{2pt}
\begin{minipage}[t]{0.48\columnwidth}
  \centering
  \includegraphics[width=\columnwidth]{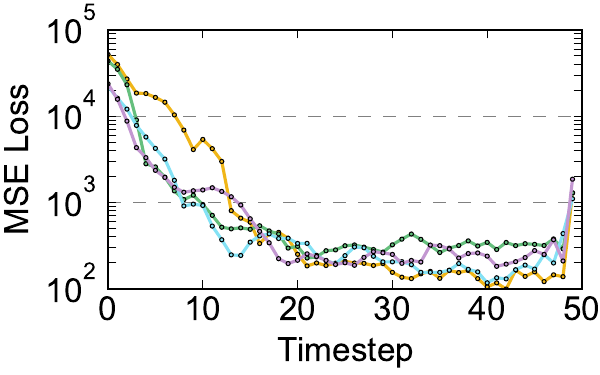}
  \caption{The sensitivity of our reuse technique accuracy to different prompts. All prompts show similar trends.}
  \label{fig:threshold_prompt}
\end{minipage}
\end{figure}


\paragraph{Idea.}
The key idea of this work is shown in \Fig{fig:idea}.
We mainly focus on self-attention. 
Specifically, our acceleration technique on attention maps consists of four steps.

\circled{white}{1}
Both the query ($Q$) and key ($K$) tokens undergo three individual checks to compute their similarity along three axes: the temporal axis, x-direction, and y-direction.
The similarity $\Delta$ is measured using the standard error metric,
\begin{equation}
\begin{aligned}
    \Delta(a) = \sqrt{\sum_{i=0}^{K-1}(a_i-\bar{a})^2/K},\quad \bar{a}=\sum_{i=0}^{K-1}a_i/K,
\end{aligned}
\end{equation}
where $a$ is a window of token channels. 
$K$ is the window size.
$\Delta(a)$ is to determine whether all channels within this window can be reused.
For example, if the window size is 2, along the temporal axis, we compute the standard error $\Delta$ between every two adjacent frames.
Token pairs with $\Delta$ below a predefined threshold $\theta_t$ are eligible for reuse in the attention computation.
The same process is applied independently along the x- and y-directions, using their respective thresholds, $\theta_x$ and $\theta_y$.

\circled{white}{2}
Once we identify reusable tokens in each dimension, we aggregate the results using a logical ``OR'' operation, i.e., a token is marked as reusable if it meets the similarity condition in any of the three axes.
\Fig{fig:idea} gives an example where specific $Q$ and $K$ tokens are marked for reuse, and reusable tokens are masked in white.
The attention computation of those tokens can be skipped, and instead, they reuse previously computed partial attention scores.

\circled{white}{3}
Next, $Q$ and $K$ perform a dot product to compute partial attention scores channel by channel.
As the white blocks of partial attention scores show in \Fig{fig:idea}, reusing saves significant computations.

\circled{white}{4}
Finally, the partial attention scores are aggregated to form the complete attention map.
Once the full attention map is obtained, the remaining operations, e.g., the Softmax operation, are applied as in the canonical self-attention.

\begin{figure}[t]
    \centering
    \begin{minipage}[t]{0.48\columnwidth}
        \centering
        \includegraphics[width=\columnwidth]{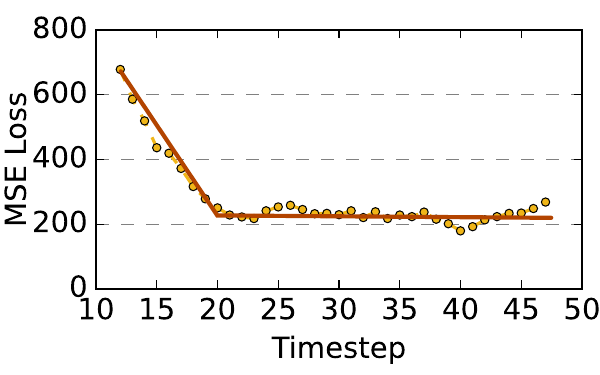}
        \subcaption{HunyuanVideo.}
    \end{minipage}
    \hspace{2pt}
    \begin{minipage}[t]{0.48\columnwidth}
        \centering
        \includegraphics[width=\columnwidth]{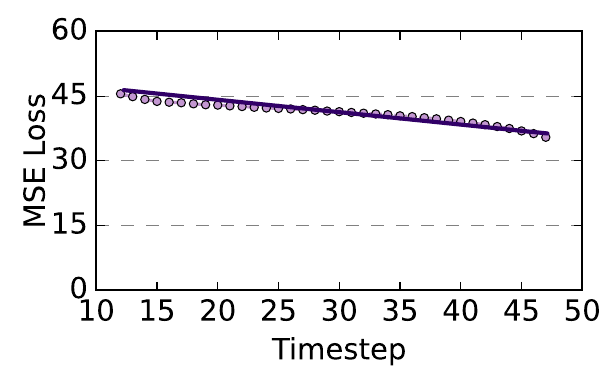}
        \subcaption{Wan2.1.}
    \end{minipage}

    \begin{minipage}[t]{0.48\columnwidth}
        \centering
        \includegraphics[width=\columnwidth]{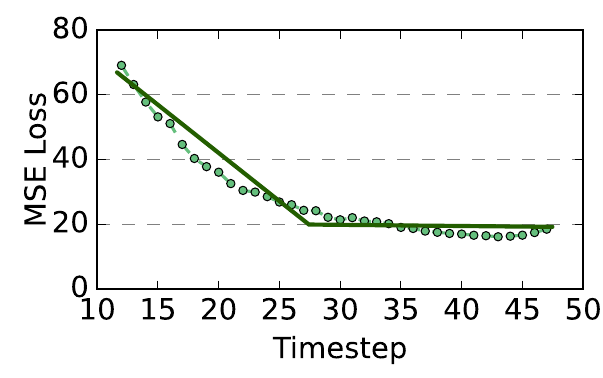}
        \subcaption{CogVideoX.}
    \end{minipage}
    \hspace{2pt}
    \begin{minipage}[t]{0.48\columnwidth}
        \centering
        \includegraphics[width=\columnwidth]{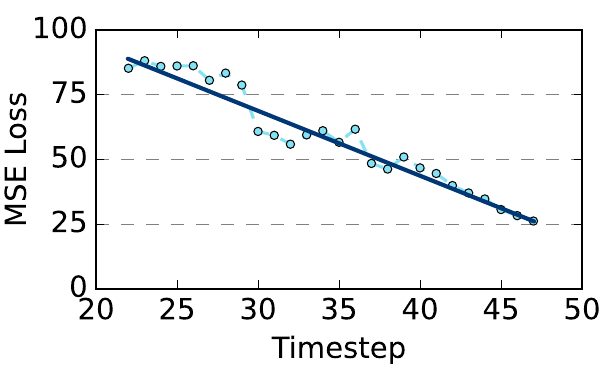}
        \subcaption{Open-Sora-Plan.}
    \end{minipage}

    \caption{The average reuse technique accuracy across the denoising steps. The straight line represents the fitted trend.}
    \label{fig:threshold_timestep}
\end{figure}

\paragraph{Why Reuse?}
\Fig{fig:reuse_cmp} explains why our proposed reuse technique is better than conventional token masking methods.
Recall, token masking exploits the insignificance of $Q$ and $K$ token values by directly skipping their related computations.
In \Fig{fig:reuse_cmp}, we evaluate our method against two masking baselines under the same token-saving ratio ($\theta^*$).
We compare the output difference between the original model and different compute-saving techniques.
The first baseline masks the top $\theta^*$ percent of token channels with the lowest values during the computation.
The second baseline uses the same selection criteria as our reuse method to identify tokens, but instead of reusing cached computation results, it directly skips their attention computations.
Results show that our technique achieves an order of magnitude lower mean-square-error (MSE) compared against these two baselines.
Intuitively, the advantage of our reuse technique is that masking techniques effectively ``reuse'' insignificant tokens that are close to the value ``0'', while our reuse strategy leverages the entire value range.

\begin{table}[t]
\centering
\caption{Hyperparameter values for each model. These values are used in our experiments with HunyuanVideo, Wan2.1, CogVideoX, and Open-Sora-Plan models.}
\resizebox{0.7\columnwidth}{!}{
\renewcommand*{\arraystretch}{1}
\renewcommand*{\tabcolsep}{3pt}
\begin{tabular}{c|cccc} 
\toprule[0.15em]
\multirow{2}{*}{Models}  & \multicolumn{4}{c}{\textbf{Hyperparameters}} \\
& $\theta_{t, \text{max}}$ & $\theta_{t, \text{min}}$ & $i_{\text{min}}$ & $i_{\text{max}}$ \\ 
\midrule[0.05em]
HunyuanVideo\cite{kong2024hunyuanvideo}           & 0.2 & 0.5 & 10 & 20 \\
Wan2.1\cite{wan2025}                 & 0.4 & 0.6 & 10 & 48 \\
CogVideoX\cite{yang2024cogvideox}            & 0.2 & 0.5 & 10 & 28 \\
Open-Sora-Plan\cite{lin2024open}        & 0.4 & 0.8 & 20 & 48 \\
\bottomrule[0.15em]
\end{tabular}
}
\label{tab:hyperparameters}
\end{table}

\paragraph{Adaptive Framework.}
While our reuse technique can achieve greater computational savings compared to prior methods, a critical question is how to determine the appropriate thresholds ($\theta_t$, $\theta_x$, and $\theta_y$) for each self-attention layer.
Our key observation is that the impact of these thresholds on the final output is highly sensitive to the denoising step, but relatively insensitive to the input prompts.
\Fig{fig:threshold_prompt} and \Fig{fig:threshold_timestep} show the sensitivity of the threshold impacts to the input prompts and the denoising step, respectively. 
We found setting $\theta_t$, $\theta_x$, and $\theta_y$ with the same threshold is more efficient and effective
(\Sect{sec:eval:sens} shows the quality comparison if we set $\theta_t$, $\theta_x$, and $\theta_y$, separately).
Our experiments show that video quality is more sensitive to threshold values during early denoising steps than in later ones.
Importantly, this trend is consistent across a range of prompts.
This allows us to share one threshold setting for all prompts.

\Fig{fig:threshold_timestep} uses HunyuanVideo as an example, which has 50 denoising steps.
We show the average MSE across 10 prompts.
The MSE monotonically decreases from denoising steps 11 to 21, while the MSE remains roughly the same from steps 22 to 49 (\Fig{fig:threshold_timestep}).
To ensure consistent impact across denoising steps, we select threshold values such that they induce the same MSE at each step.
Specifically, the threshold $\theta_{t, i}$ at denoising step $i$ is defined as:
\begin{equation}
    \theta_{t, i} =  (i - i_{\text{min}}) \cdot \frac{\theta_{t, \text{max}} - \theta_{t, \text{min}}}{i_{\text{max}} - i_{\text{min}}}, \ \text{where} \ i \in [i_{\text{min}}, i_{\text{max}}], \\
\end{equation}
where $\theta_{t, \text{max}}$ and $\theta_{t, \text{min}}$ are two hyper-parameters.
The values $i_{\text{min}} = 11$ and $i_{\text{max}} = 21$ define the range over which the threshold linearly increases.
For steps outside this range, we keep the first 10 steps and the last step untouched.
For the remaining steps, we apply a fixed threshold of $\theta_{t, \text{max}}$. 
These parameters are summarized in \Tbl{tab:hyperparameters}.

\section{Evaluation}
\label{sec:eva}

\begin{table*} 
\caption{Quantitative evaluation of our method, \proj, against the state-of-the-arts~\cite{chen2024delta, zhao2024real, jiang2024minference, xi2025sparse} on four widely-adopted vDiTs: HunyuanVideo~\cite{kong2024hunyuanvideo}, CogVideoX~\cite{yang2024cogvideox} and Open-Sora-Plan~\cite{lin2024open}. We annotate the \goldtext{best} and \silvertext{second-best} results among all methods.}
\resizebox{\textwidth}{!}{
\renewcommand*{\arraystretch}{1}
\begin{tabular}{c|c|cccc|ccccc} 
\toprule[0.15em]
\multirow{2}{*}{Model}  & \multirow{2}{*}{Methods} & \multicolumn{4}{c|}{\textbf{Quality Metrics}} & \multicolumn{5}{c}{\textbf{Performance Metrics}}  \\
& & \specialcell{VBench$\uparrow$\\(\%)} & \specialcell{PSNR $\uparrow$\\(dB)} & SSIM$\uparrow$ & LPIPS$\downarrow$ &  \specialcell{Latency $\downarrow$\\(s)} & Speedup$\uparrow$ & \specialcell{Theoretical\\Speedup $\uparrow$} & \specialcell{FLOPs$\downarrow$\\($10^{15}$)}  & \specialcell{Mem.$\downarrow$\\(GB)} \\ 
\midrule[0.05em]
\multirow{9}{*}{HunyuanVideo~\cite{kong2024hunyuanvideo}} & \cellcolor{blue!8} Original & \cellcolor{blue!8} 80.28 & \cellcolor{blue!8} - & \cellcolor{blue!8} - & \cellcolor{blue!8} -  & \cellcolor{blue!8} 694.03 & \cellcolor{blue!8} - & \cellcolor{blue!8} - & \cellcolor{blue!8} 214.27 & \cellcolor{blue!8} 45.81 \\
& \mode{$\Delta$-DiT}~\cite{chen2024delta} & \silver\silvertext{80.43} & 26.09 & 0.862 & 0.111 & 569.43 & 1.22 & 1.53 & 145.09 & 46.59 \\
& \mode{PAB$_{5,9}$}~\cite{zhao2024real} & 78.90 & 26.07 & 0.778 & 0.084 & 562.92 & 1.23 & 1.61 & 145.55 & 57.58 \\
& \mode{MInference}~\cite{jiang2024minference} & 80.18 & 25.00 & 0.825 & 0.174 & 481.18 & 1.44 & 2.02 & 125.22 & 50.88 \\
& \mode{SVG$_{70\%}$}~\cite{xi2025sparse} & 79.97 & 25.78 & 0.843 & 0.153 & 409.57 & 1.69 & 2.12 & 121.51 & 50.87 \\
& \mode{\proj$_{85\%}$} & \gold\goldtext{80.44} & \silver\silvertext{31.28} & \silver\silvertext{0.915} & \silver\silvertext{0.078} & \gold\goldtext{260.85} & \gold\goldtext{2.66} & \gold\goldtext{2.66} & \gold\goldtext{105.52} & \gold\goldtext{45.81} \\
& \mode{\proj$_{75\%}$} & {80.23} & \gold\goldtext{35.06} & \gold\goldtext{0.950} & \gold\goldtext{0.036} & 329.97 & 2.10 & 2.10 & 122.02 & \gold\goldtext{45.81} \\
& \mode{\proj$_{75\%}$+SVG$_{70\%}$} & 79.84 & 26.17 & 0.851 & 0.142 & \silver\silvertext{285.86} & \silver\silvertext{2.43} & \silver\silvertext{2.43} & \silver\silvertext{111.46} & 50.87 \\
\midrule[0.05em]
\multirow{6}{*}{Wan2.1~\cite{wan2025}} & \cellcolor{blue!8} Original & \cellcolor{blue!8} 81.17 & \cellcolor{blue!8} - & \cellcolor{blue!8} - & \cellcolor{blue!8} - & \cellcolor{blue!8} 528.43 & \cellcolor{blue!8} - & \cellcolor{blue!8} - & \cellcolor{blue!8} 117.94 & \cellcolor{blue!8} 46.66 \\
& \mode{$\Delta$-DiT}~\cite{chen2024delta} & 80.54 & 16.33 & 0.531 & 0.393 & 462.74 & 1.14 & 1.28 & 144.90 & \silver\silvertext{46.97} \\
& \mode{PAB$_{5,9}$}~\cite{zhao2024real} & 78.94 & \silver\silvertext{22.84} & 0.698 & 0.185 & \silver\silvertext{254.09} & \silver\silvertext{2.08} & \gold\goldtext{2.36} & \gold\goldtext{105.55} & 71.66 \\
& \mode{MInference}~\cite{jiang2024minference} & \silver\silvertext{81.05} & 18.33 & 0.632 & 0.253 & 379.93 & 1.39 & 1.89 & 122.74 & 47.27 \\
& \mode{SVG$_{70\%}$}~\cite{xi2025sparse} & \gold\goldtext{81.47} & 22.83 & \silver\silvertext{0.778} & \silver\silvertext{0.142} & 382.05 & 1.38 & 1.89 & 122.74& 47.27 \\
& \mode{\proj$_{80\%}$} & 81.03 & \gold\goldtext{27.00} & \gold\goldtext{0.862} & \gold\goldtext{0.070} & \gold\goldtext{225.70} & \gold\goldtext{2.34} & \silver\silvertext{2.34} & \silver\silvertext{111.25} & \gold\goldtext{46.66} \\
\midrule[0.05em]
\multirow{6}{*}{CogVideoX~\cite{yang2024cogvideox}} & \cellcolor{blue!8} Original & \cellcolor{blue!8} 80.95 & \cellcolor{blue!8} - & \cellcolor{blue!8} - & \cellcolor{blue!8} - & \cellcolor{blue!8} 772.25 & \cellcolor{blue!8} - & \cellcolor{blue!8} - & \cellcolor{blue!8} 117.94 & \cellcolor{blue!8} 23.51 \\
& \mode{$\Delta$-DiT}~\cite{chen2024delta} & 77.46 & 20.34 & 0.723 & 0.341 & 695.48 & 1.11 & 1.15 & 104.64 & \silver\silvertext{24.03} \\
& \mode{PAB$_{5,9}$}~\cite{zhao2024real} & 68.17 & 11.30 & 0.486 & 0.716 & \silver\silvertext{464.42} & \silver\silvertext{1.66} & \silver\silvertext{2.07} & \silver\silvertext{72.94} & 28.55 \\
& \mode{MInference}~\cite{jiang2024minference} & 74.73 & 14.63 & 0.575 & 0.524 & 611.55 & 1.26 & 1.69 & 80.86 & 42.68 \\
& \mode{SVG$_{70\%}$}~\cite{xi2025sparse} & \silver\silvertext{80.87} & \silver\silvertext{23.19} & \silver\silvertext{0.799} & \silver\silvertext{0.202} & 590.64 & 1.31 & 1.75 & 79.37 & 42.68 \\
& \mode{\proj$_{80\%}$} & \gold\goldtext{81.16} & \gold\goldtext{25.58} & \gold\goldtext{0.846} & \gold\goldtext{0.134} & \gold\goldtext{333.95} & \gold\goldtext{2.31} & \gold\goldtext{2.31} & \gold\goldtext{69.27} & \gold\goldtext{23.51} \\
\midrule[0.05em]
\multirow{6}{*}{Open-Sora-Plan~\cite{lin2024open}} & \cellcolor{blue!8} Original & \cellcolor{blue!8}72.44 & \cellcolor{blue!8} - & \cellcolor{blue!8} - & \cellcolor{blue!8} - & \cellcolor{blue!8} 44.82 & \cellcolor{blue!8} - & \cellcolor{blue!8} - & \cellcolor{blue!8} 14.93 & \cellcolor{blue!8} 35.91 \\
& \mode{$\Delta$-DiT}~\cite{chen2024delta} & 70.21 & \silver\silvertext{16.30} & \silver\silvertext{0.631} & \silver\silvertext{0.306} & 35.96 & 1.25 & 1.28 & 11.65 & 36.70 \\
& \mode{PAB$_{5,9}$}~\cite{zhao2024real} & 72.61 & 13.51 & 0.520 & 0.403 & \gold\goldtext{27.28} & \gold\goldtext{1.64} & \gold\goldtext{1.66} & \gold\goldtext{8.99} & 41.20 \\
& \mode{MInference}~\cite{jiang2024minference} & 70.84 & 15.38 & 0.598 & 0.338 & 41.49 & 1.08 & 1.22 & 12.22 & \silver\silvertext{36.61} \\
& \mode{SVG$_{70\%}$}~\cite{xi2025sparse} & \silver\silvertext{73.01} & 15.32 & 0.591 & 0.346 & 39.51 & 1.13 & 1.27 & 11.74 & \silver\silvertext{36.61} \\
& \mode{\proj$_{80\%}$} & \gold\goldtext{73.18} & \gold\goldtext{18.55} & \gold\goldtext{0.706} & \gold\goldtext{0.226} & \silver\silvertext{31.08} & \silver\silvertext{1.44} &  \silver\silvertext{1.44} &  \silver\silvertext{10.35} & \gold\goldtext{35.91} \\
\bottomrule[0.05em]
\end{tabular}
}
\label{tab:overall}
\end{table*}

\subsection{Experimental Setup}
\label{sec:eva:exp}

\paragraph{Baselines.} 
We evaluate our technique on four vDiT models: HunyuanVideo-T2V~\cite{kong2024hunyuanvideo}, Wan2.1-T2V-14B\cite{wan2025}, CogVideoX1.5-5B-T2V~\cite{yang2024cogvideox} and Open-Sora-Plan\cite{lin2024open}.
For HunyuanVideo-T2V, we generate 5.33-second videos with a resolution of $544 \times 960$.
For Wan2.1-T2V-14B, we generate 5-second videos with a resolution of $480 \times 832$.
For CogVideoX1.5-5B-T2V, we generate 5-second videos with a resolution of $768 \times 1360$.
For Open-Sora-Plan, we generate 1.2-second videos with a resolution of $640 \times 480$
Our evaluation compares with two the state-of-the-art techniques for attention accelerations: \mode{MInference}~\cite{jiang2024minference}, \mode{Sparse Videogen} (\mode{SVG})~\cite{xi2025sparse}.
We also compare with acceleration methods: \mode{PAB}~\cite{zhao2024real} and \mode{$\Delta$-DiT}~\cite{chen2024delta}.

\paragraph{Metrics and Hardware.} 
Following prior works~\cite{zhao2024real, zou2024accelerating}, we use the VBench~\cite{huang2024vbench} as the video quality metric.
During the experiments, we generate 5 videos for each of the 950 benchmark prompts using different random seeds and then average the VBench scores.
The generated videos are evaluated across 16 aspects from VBench.
We report the average score on VBench.
For image quality evaluation, we use PSNR, SSIM, and LPIPS.
We compare the generated videos by different acceleration methods against the original video results frame-by-frame.
For performance, we report end-to-end generation delay, GPU memory consumption, and computational complexity (FLOPs).
All performance are measured on a single Nvidia H100 (80 GB).

\begin{figure*}[!t]
    \centering
    \includegraphics[width=0.95\textwidth]{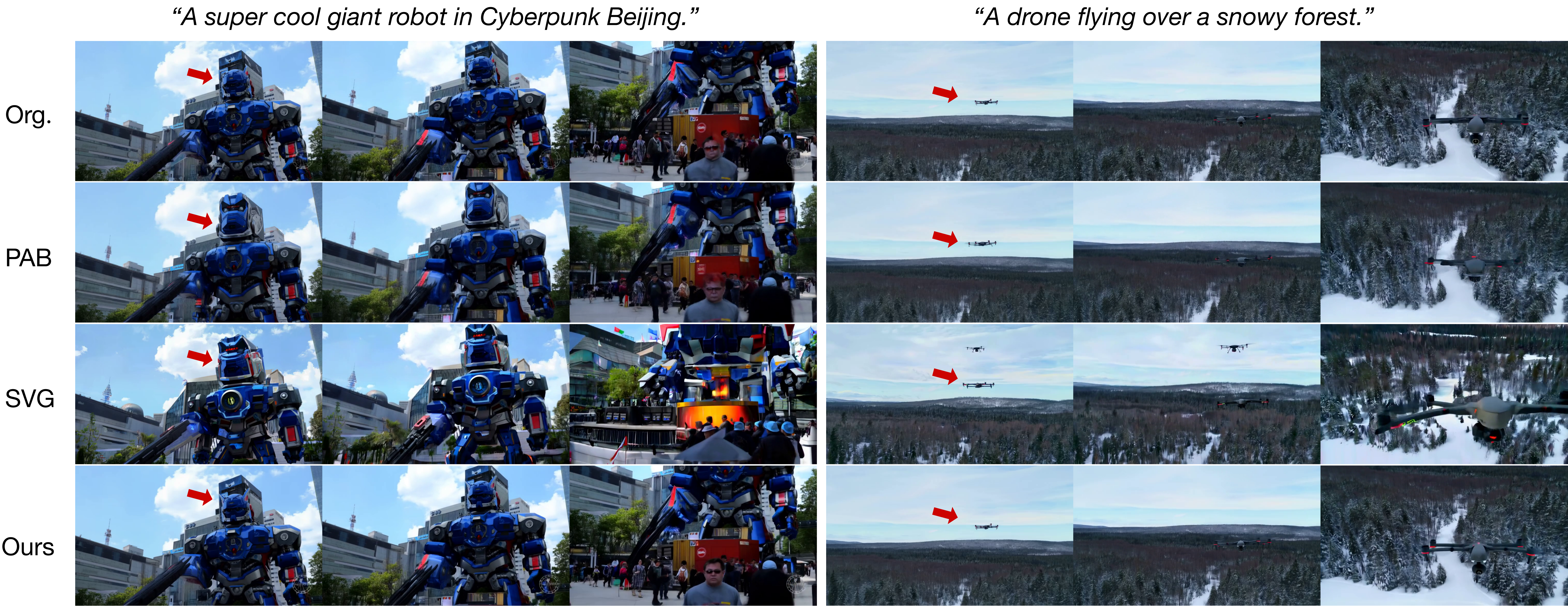}
    \caption{The qualitative evaluations of \proj against other methods. ``\redright'' shows the major artifacts in prior studies. 
    }
    \label{fig:examples}
\end{figure*}

\subsection{Accuracy and Performance}
\label{sec:eva:perf}

\paragraph{Configurations.} 
\Tbl{tab:overall} compares \proj with state-of-the-art methods. 
\mode{$\Delta$-DiT} performs coarse block-wise reuse. 
\mode{PAB$_{5,9}$} reuse attention results in the middle 80\% of denoising steps, with window sizes of (5, 9) for self-attention and cross-attention.
\mode{MInference} achieves 52\% self-attention reduction via predefined patterns.
\mode{SVG$_{70\%}$} retains full computation in the first 12 steps, and sparsifies the rest to 30\%.
We evaluate two variants of our method: \mode{\proj$_{75\%}$}, which adaptively skips roughly 75\% of attention computations from step 11 onward, and \mode{\proj$_{75\%}$+SVG$_{70\%}$}, which combines our reuse strategy with sparse masking.

\paragraph{Video Quality.}
\proj achieves almost the best VBench scores across all models and baselines. 
On HunyuanVideo, \mode{\proj$_{85\%}$} surpasses the original model, possibly due to regularization effects. 
However, it underperforms on other metrics against \mode{\proj$_{75\%}$}, which offers the best trade-off between quality and acceleration.

\paragraph{Image Quality.} 
Across PSNR, SSIM, and LPIPS, \proj consistently outperforms prior methods.
On HunyuanVideo, \proj achieves 35.1~dB PSNR vs. 26.09~dB from the best baseline (\mode{$\Delta$-DiT}), outperforming all others by large margins.
Similar gains are observed on the other two models, indicating strong generalization.
\Fig{fig:examples} illustrates qualitative results, showing \proj retains finer details compared to other methods.

\paragraph{Performance.} 
\proj also achieves the highest speedup among all evaluated baselines.
Note that, existing self-attention acceleration techniques, such as FlashAttention~\cite{dao2022flashattention, dao2023flashattention}, do not support our reuse strategy.
Therefore, we estimate speedup by proportionally reducing the self-attention latency based on the amount of computation reduced.
On HunyuanVideo, \mode{\proj$_{75\%}$} achieves a theoretical speedup of 2.1$\times$ compared to the original model, whereas \mode{SVG$_{70\%}$} and \mode{MInference} achieve only 1.7$\times$ and 1.4$\times$, respectively.
\mode{\proj$_{85\%}$} can even boost the performance to 2.7$\times$.

We further compare the reduction in the number of operations during the entire execution.
On average, \proj reduces computation by 75\%, which is 5\% higher than \mode{SVG$_{70\%}$}.
Meanwhile, our method introduces only minimal runtime memory overhead, making it more efficient and scalable than prior approaches.
For instance, \mode{PAB$_{5, 9}$} requires caching intermediate results, while we only reuse within the current self-attention.
Additionally, \mode{\proj$_{75\%}$+SVG$_{70\%}$} combines our technique with SVG.
This variant further boosts the performance to 2.4$\times$ speedup while introducing only a minimal accuracy loss.

\subsection{Ablation Study}

\begin{table}[t]
\centering
\caption{Ablation study of our method. Here, we only show the result with HunyuanVideo~\cite{kong2024hunyuanvideo}. 
}
\resizebox{\columnwidth}{!}{
\renewcommand*{\arraystretch}{1}
\renewcommand*{\tabcolsep}{3pt}
\begin{tabular}{c|cccc} 
\toprule[0.15em]
\multirow{2}{*}{Methods} & \multicolumn{4}{c}{\textbf{Quality Metrics}} \\
& VBench$\uparrow$ & PSNR (dB)$\uparrow$ & SSIM$\uparrow$ & LPIPS$\downarrow$ \\ 
\midrule[0.05em]
\rowcolor{blue!8}
Original           & 80.28 & - & - & - \\
\mode{\proj$_{75\%}$} (Fixed) & 80.19 & 34.56 & 0.948 & 0.037  \\
\mode{\proj$_{75\%}$} (Temp) & 80.13 & 32.44 & 0.933 & 0.050 \\
\mode{\proj$_{75\%}$} (Spat+Temp) & 80.23 & 35.06 & 0.950 & 0.036 \\
\bottomrule[0.15em]
\end{tabular}
}
\label{tab:ablation}
\end{table}


\Tbl{tab:ablation} presents the contributions of individual components in \proj through an ablation study.
Here, we evaluate two additional variants.
The first variant uses a fixed threshold across all denoising steps.
The second variant adopts adaptive thresholds but applies reuse only along the temporal dimension without spatial correlations.
For a fair comparison, all variants are configured to achieve roughly the same level of computational savings.
The percentage shows the savings on self-attention.

Our variant with a fixed threshold still achieves a VBench score of 80.19\%, which is significantly higher than the scores of other baselines reported in \Tbl{tab:overall}.
However, our adaptive thresholding strategy further improves performance, as it better aligns with the MSE sensitivity trend in \Fig{fig:threshold_timestep}.
We also show the quality result of a variant that only uses temporal correlation.
Thus, using one dimension does not unleash the full potential of our reuse technique.

\subsection{Sensitivity Study}
\label{sec:eval:sens}

\begin{figure}[t]
  \centering
  \includegraphics[width=0.3\textwidth]{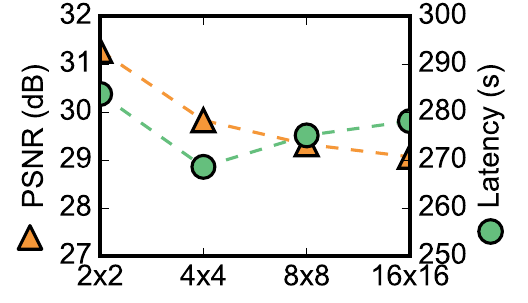}
  \caption{Sensitivity of our method to the reuse window size.}
  \label{fig:sens}
\end{figure}

\begin{table}[t]
\centering
\caption{Sensitivity study of generation quality to dynamic thresholds. Here, we only show the result with HunyuanVideo~\cite{kong2024hunyuanvideo}. Results with other models show a similar trend.}
\resizebox{\columnwidth}{!}{
\renewcommand*{\arraystretch}{1}
\renewcommand*{\tabcolsep}{3pt}
\begin{tabular}{c|cccc} 
\toprule[0.15em]
\multirow{2}{*}{Methods} & \multicolumn{4}{c}{\textbf{Quality Metrics}} \\
& VBench$\uparrow$ & PSNR (dB)$\uparrow$ & SSIM$\uparrow$ & LPIPS$\downarrow$ \\ 
\midrule[0.05em]
\mode{\proj$_{85\%}$} & 80.44 & 31.28 & 0.915 & 0.078  \\
\mode{\proj$_{85\%}$} (channel wise) & 79.37 & 28.27 & 0.912 & 0.074\\
\bottomrule[0.15em]
\end{tabular}
}
\label{tab:sensitivity}
\end{table}

\Fig{fig:sens} shows the sensitivity of our method to different reuse window sizes on HunyuanVideo.
We use the same reuse threshold as \mode{\proj$_{85\%}$}.
In all experiments, the reuse threshold is set to the same as that used for a window size of 2.
Our results show that increasing the window size to 4 leads to a noticeable degradation in image quality.
The reduction in image quality is from excessive reuse of tokens.
Meanwhile, increasing the reuse window size over 4 leads to a performance drop. The drop in speedup is caused by a decrease in the number of tokens that meet the reuse condition at larger window sizes.
These findings suggest that a window size of 2 provides a good trade-off.

\Tbl{tab:sensitivity} shows the sensitivity analysis of our method with individually setting thresholds for the channel dimension. 
The threshold $\tau$ is defined as $
\tau = \alpha \cdot \frac{1}{D}\sum_{i=1}^{D} \lvert \Delta_i \rvert, $
where $\alpha$ is a base coefficient and $\Delta_i$ denotes the channel-wise difference.
This adaptive formulation stabilizes the reuse rate across channels by normalizing the threshold with respect to the mean absolute variation along the last dimension.
The actual threshold setting and detailed experimental setup are shown in the supplementary.
The result shows a slight decrease in video quality. 
We believe this quality degradation is the attention score is the sum of all channels' partial results.
Thus, it is better to set their thresholds uniformly.
\section{Discussion and Conclusion}
\label{sec:conc}

As the field of video generation rapidly advances, the need for efficient frameworks has become increasingly urgent.
This work is the first to identify the inherent spatio-temporal correlations across tokens in the latent space of vDiTs.
By understanding such correlations, we propose a novel reuse-based strategy that significantly reduces redundant self-attention computations.
By evaluating across the mainstream vDiTs, we show that our approach achieves up to 85\% computational savings on self-attention while preserving near-identical quality on generated videos.

\paragraph{Limitation.} 
Nevertheless, our method introduces unstructured sparsity in the attention computation, which may limit compatibility with existing low-level self-attention acceleration frameworks~\cite{dao2022flashattention, dao2023flashattention, ye2025flashinfer, zheng2023efficiently, kwon2023efficient}.
How to integrate our reuse strategy with such frameworks remains an important direction for our future work.

{
    \small
    \bibliographystyle{ieeenat_fullname}
    \bibliography{ref}
}


\end{document}